\documentclass[showpacs,aps,twocolumn,superscriptaddress,prb,amsmath]
{revtex4}
\usepackage{txfonts}
\usepackage{color}
\usepackage{amssymb}
\usepackage{graphicx}
\usepackage{dcolumn}
\usepackage{bm}
\usepackage{hyperref}
\usepackage{color}

\begin{document}
\preprint{APS/123-QED}
\title{Complex low energy tetrahedral polymorphs of group IV elements from first-principles}
\author{Chaoyu He}
\email{hechaoyu@xtu.edu.cn}\affiliation{Hunan Key Laboratory for Micro-Nano Energy Materials and Devices, Xiangtan University, Hunan 411105, P. R. China;}
\affiliation{School of Physics and Optoelectronics, Xiangtan University, Xiangtan 411105, China.}
\author{Xizhi Shi}
\affiliation{Hunan Key Laboratory for Micro-Nano Energy Materials and Devices, Xiangtan University, Hunan 411105, P. R. China;}
\affiliation{School of Physics and Optoelectronics, Xiangtan University, Xiangtan 411105, China.}
\author{S. J. Clark} 
\affiliation{Durham University, Centre for Material Physics, Department of Physics, South Road, Durham, DH1 3LE, United Kingdom}
\author{Jin Li}
\affiliation{Hunan Key Laboratory for Micro-Nano Energy Materials and Devices, Xiangtan University, Hunan 411105, P. R. China;}
\affiliation{School of Physics and Optoelectronics, Xiangtan University, Xiangtan 411105, China.}
\author{Chris J. Pickard}
\email{cjp20@cam.ac.uk}
\affiliation{Department of Materials Science $\&$ Metallurgy, University of Cambridge, 27 Charles Babbage Road, Cambridge CB3$~$0FS, United Kingdom}
\affiliation{Advanced Institute for Materials Research, Tohoku University 2-1-1 Katahira, Aoba, Sendai, 980-8577, Japan}
\author{Tao Ouyang}
\affiliation{Hunan Key Laboratory for Micro-Nano Energy Materials and Devices, Xiangtan University, Hunan 411105, P. R. China;}
\affiliation{School of Physics and Optoelectronics, Xiangtan University, Xiangtan 411105, China.}
\author{Chunxiao Zhang}
\affiliation{Hunan Key Laboratory for Micro-Nano Energy Materials and Devices, Xiangtan University, Hunan 411105, P. R. China;}
\affiliation{School of Physics and Optoelectronics, Xiangtan University, Xiangtan 411105, China.}
\author{Chao Tang}  
\affiliation{Hunan Key Laboratory for Micro-Nano Energy Materials and Devices, Xiangtan University, Hunan 411105, P. R. China;}
\affiliation{School of Physics and Optoelectronics, Xiangtan University, Xiangtan 411105, China.}
\author{Jianxin Zhong}
\affiliation{Hunan Key Laboratory for Micro-Nano Energy Materials and Devices, Xiangtan University, Hunan 411105, P. R. China;}
\affiliation{School of Physics and Optoelectronics, Xiangtan University, Xiangtan 411105, China.}
\date{\today}
\begin{abstract}
The energy landscape of carbon is exceedingly complex, hosting diverse and important metastable phases, including diamond, fullerenes, nanotubes and graphene. Searching for structures, especially those with large unit cells, in this landscape is challenging. Here we use a combined stochastic search strategy employing two algorithms (AIRSS and RG$^2$) to apply connectivity constraints to unit cells containing up to 100 carbon atoms. We uncover three low energy carbon polymorphs (Pbam-32, P6/mmm and I$\overline{4}$3d) with new topologies, containing 32, 36 and 94 atoms in their primitive cells, respectively. Their energies relative to diamond are 96 meV/atom, 131 meV/atom and 112 meV/atom, respectively, which suggests potential metastability. These three carbon allotropes are mechanically and dynamically stable, insulating carbon crystals with superhard mechanical properties. The I$\overline{4}$3d structure possesses a direct band gap of 7.25 eV, which is the widest gap in the carbon allotrope family. Silicon, germanium and tin versions of Pbam-32, P6/mmm and I$\overline{4}$3d also show energetic, dynamical and mechanical stability. The computed electronic properties show that they are potential materials for semiconductor and photovoltaic applications.
\end{abstract}
\maketitle


The group IV elements carbon and silicon are central to life and technology respectively. Carbon can form many allotropes in addition to graphite and cubic diamond (cd). For example, quasi zero-dimensional fullerenes, quasi one-dimensional nanotubes, quasi two-dimensional graphene and other three-dimensional crystals such as hexagonal diamond (hd), chaoite \cite{1, 2, 3} and amorphous carbon. In the past decades, hypothetical crystalline structures of carbon \cite{4, 5} have been predicted. Some of these structures, such as M-carbon \cite{6, 7}, S-carbon \cite{8, 9}, C20-sc, C21-sc and C22-sc\cite{10, 11} are believed to have been synthesized in previous experiments. \cite{12, 13} Others (such as P6522 \cite{14}, P42/ncm \cite{15}, Pbam-24 \cite{16}, calt34 \cite{17} and calt46 \cite{17}) also possess better energetic stabilities in comparison with cd and are targets for future experimental synthesis. Some have been reported as experimentally realized crystals, for example, the monoclinic V-carbon \cite{18} was obtained by compressing C70 pea-pods and the previously predicted T-carbon \cite{19} was recently claimed to have been synthesized through pseudo-topotactic conversion of a multi-walled carbon nanotube \cite{20}.

\begin{figure}
\center
\includegraphics[width=\columnwidth]{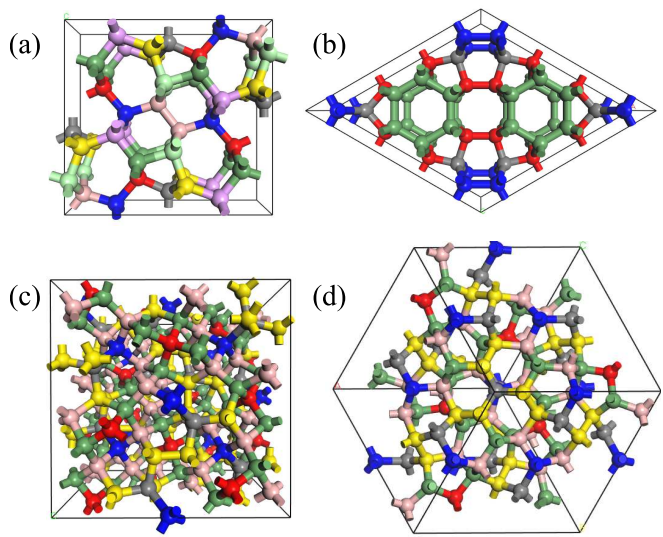}
\caption{The perspective crystalline views of the optimized carbon-version Pbam-32 (a) and P6/mmm (b), as well as those of I$\overline{4}$3d in both crystalline cell (c) and primitive cell (d). Balls in different colors indicate different inequivalent carbon atoms.}\label{fig1}
\end{figure}

Silicon, germanium and tin adopt similar structures to carbon, and have also attracted many theoretical efforts to explore their polymorphs \cite{21, 22, 15,23, 16, 24, 25, 26, 27, 28, 29, 30}. Advances accessing silicon polymorphs through unconventional pathways, such as ultrafast laser-induced confined micro-explosion \cite{31} and high-pressure precursor routes \cite{32}, has increased interest in silicon and germanium \cite{33}. Low energy silicon allotropes with direct-band gaps of the appropriate size are anticipated for future applications to solar-energy conversion \cite{34, 29, 30, 28}.

The development of rapid and reliable codes for the computation of material properties, combined with the advent of algorithms for the prediction of crystal structures \cite{35, 36, 37, 38} and an increase in the performance and availability of computational resources allows us to discover new materials from first principles. In this paper, we report three low energy tetrahedral polymorphs (identified by their space groups, Pbam-32, P6/mmm and I$\overline{4}$3d) for carbon, silicon, germanium and tin, which are obtained using a stochastic search strategy as implemented in the AIRSS \cite{35, 36} and the RG$^2$-codes \cite{5}. Density functional theory (DFT) based first-principles calculations as implemented in the Vienna ab initio simulation package (VASP) \cite{39} and CASTEP \cite{40} were employed to investigated their structures, energetic, dynamical and mechanical stabilities, as well as electronic and mechanical properties. Our results show that these three tetrahedral polymorphs possess remarkable energetic stabilities in carbon, silicon, germanium and tin, in comparison with those metastable phases previously proposed. They are confirmed to be both dynamically and mechanically stable. The carbon Pbam-32, P6/mm and I$\overline{4}$3d structures are superhard insulators with excellent mechanical properties. In silicon, they are potential absorber materials for thin-film solar-cell applications according to their proper band gaps. Notably, the carbon I$\overline{4}$3d structure is a wide-gap semiconductor with a 7.25 $eV$ gap, which is believed as the widest-gap in carbon family.

\begin{figure}
\center
\includegraphics[width=\columnwidth]{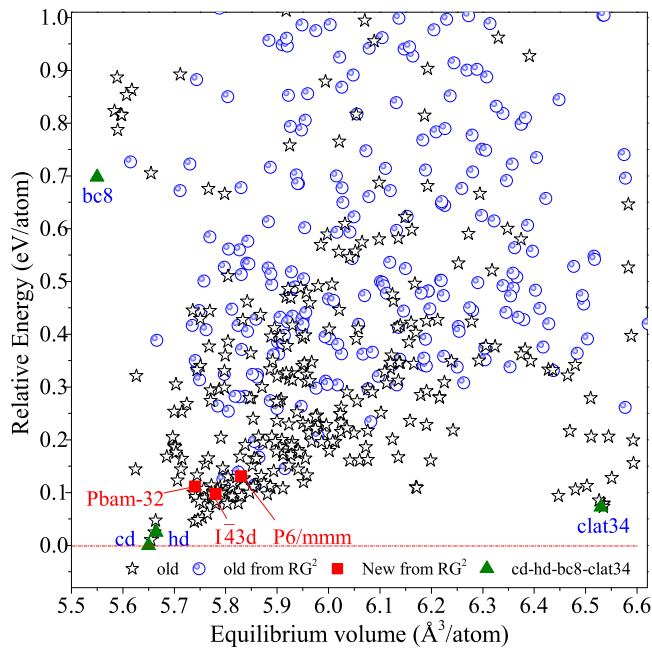}
\caption{Scatter plot of the PBE-based relative average energy (eV/atom) against equilibrium volume ($\AA^3$) of the carbon-version Pbam-32, P6/mmm, I$\overline{4}$3d predicted in this work (red squares), the sp$^3$ carbons previously predicted in the past decades \cite{4} (black five-point stars), as well as those sp$^3$ carbons discovered in our recent stochastic search \cite{5} (blue circles). Some well-known carbon phases (green triangles)are also plotted in for comparison.}\label{fig2}
\end{figure}

\begin{table*}
  \caption{The PBE-based total energies (E$_{tot}$: meV/atom) relative to the corresponding cd form, equilibrium volume per atom (V$_0$: ${\AA}^3$) and the energy band gaps (indirect/direct) calculated based on PBE (E$_g^{PBE}$: eV) and HSE06 (E$_g^{HSE}$: eV) for various typical phases of carbon, silicon, germanium and tin.}
\begin{tabular}{c c c c c c c c c c c c c c c c c}
\hline \hline
            &    &      &C       &         &   &      &Si        &         &    &      &Ge       &         &   &      &Sn       &         \\
System &E$_{tot}$&$V_0$&E$_g^{PBE}$&E$_g^{HSE}$&E$_{tot}$&$V_0$&E$_g^{PBE}$&E$_g^{HSE}$&E$_{tot}$&$V_0$ &E$_g^{PBE}$&E$_g^{HSE}$ &E$_{tot}$ &$V_0$ &E$_g^{PBE}$ &E$_g^{HSE}$ \\
\hline
Pbam-32      &98  &5.78 &4.76/4.89&5.97/6.22&33 &20.71 &0.81      &1.39     &31  &24.52 &0.08     &0.66     &26 &37.33 &Metal    &0.07     \\
P6/mmm      &131 &5.83 &3.55/4.75&4.76/5.93&46 &20.96 &0.56      &1.13     &46  &24.76 &0.27     &0.77/0.88&39 &37.68 &Metal    &0.40/0.46\\
I$\overline{4}$3d   &112 &5.74 &5.91     &7.24     &39 &20.44 &1.54/1.61 &2.19/2.26&35  &24.18 &0.85/1.06&1.41/1.58&31 &36.74&0.47/0.64&0.86/1.03\\
\hline
cd          &0   &5.65 &4.15/5.59&5.32/6.99&0  &20.44 &0.62/2.56 &1.19/3.33&0   &24.13 &Metal    &0.14     &0  &36.83 &Metal    &Metal    \\
Pbam-24     &68  &5.81 &4.56/4.79&-        &29 &20.97 &0.85/0.87 &-        &27  &24.83 &Metal    &-        &24 &37.81 &Metal    &-        \\
clat34      &73  &6.53 &3.74/3.77&-        &51 &23.54 &1.38      &-        &21  &27.49 &0.59/0.59&-        &19 &41.76 &0.33     &-        \\
V-carbon    &105 &5.79 &4.44/4.76&-        &48 &20.65 &0.85/1.03 &-        &45  &24.52 &Metal    &-        &35 &37.01 &Metal    &-        \\
clat46      &106 &6.48 &3.85     &-        &62 &23.25 &1.31/1.32 &-        &27  &27.21 &1.03/1.06&-        &23 &41.29 &0.66/0.67&-        \\
I-4(I1373)  &161 &6.06 &5.25/5.26&6.55/6.56&62 &21.72 &1.29/1.66 &1.91/2.30&52  &25.59 &1.15/1.21&1.55/1.78&46 &38.90 &0.73/0.78&1.07/1.19\\
Si24        &235 &6.18 &2.92/3.60&4.06/4.75&92 &21.99 &0.51/0.58 &1.07/1.13&85  &26.05 &0.09/0.22&0.59/0.71&74 &32.04 &0.12/0.25&0.51/0.64\\
tp12(st12)  &886 &5.59 &4.98/5.49&6.27/6.89&166&18.33 &1.11/1.39 &1.71/1.96&136 &21.72 &0.25     &0.69     &60 &31.36 &Metal    &Metal    \\
T-carbon    &1173&13.18&2.287    &-        &1024&52.16&Metal     &-        &943 &68.41 &Metal    &-        &792&109.75&Metal    &-        \\
\hline
\end{tabular}
\label{tab-1}
\end{table*}

Perspective views of the carbon Pbam-32, P6/mmm and I$\overline{4}$3d structures are shown in Fig.~\ref{fig1}. The corresponding structural information is provided in Tab.S3 in the supplemental material \cite{39}. The Pbam-32 structure is structurally very similar to the previously proposed Pbam-24 structure\cite{16} with 5-fold, 6-fold and 7-fold topological carbon-rings. Pbam-32 and Pbam-24 are orthorhombic phases belonging to Pbam (No.55) symmetry with 8/32 and 6/24 inequivalent/total carbon atoms in their unit cell, respectively. We find that both Pbam-24 and Pbam-32 have straightforward transformation paths from graphite (see Fig.S1 \cite{39}) and can be considered as good candidates for explaining the diversity of products on the cold compression of graphite \cite{12}. Their simulated X-ray diffraction patterns (XRD) are provided and compared with the experimental results Fig.S3 \cite{39}, which show that both Pbam-32 and Pbam-24 can partially explain the experimental XRD.

The cold compression of carbon nanotube (CNT) bundles \cite{41} is another route to synthesize novel carbon phases, and is an effective theoretical way to design new carbon crystals\cite{42, 43}. The corresponding transition pathways from suitably arranged (4, 0)-CNTs to Pbam-32 and Pbam-24 are suggested in Fig.S1 \cite{39}. These discoveries indicate that both Pbam-24 and Pbam-32 may be synthesized by cold compressing graphite \cite{12} and CNTs \cite{41}, similar to the V-carbon (105 meV/atom) recently synthesized by compressing C70 pea-pods \cite{18}.

The P6/mmm structure possesses relatively high symmetry (No. 191) in comparison with Pbam-32. There are 36 carbon atoms in the crystalline cell of P6/mmm and only 4 of them are inequivalent. The P6/mmm structure contains 5-fold, 6-fold and 8-fold topological carbon-rings, which is similar to the previously proposed low-energy M585 phase \cite{44}. A simple pathway for structurally transforming perfect graphite to P6/mmm is currently unknown. However, it can be structurally transformed from the well-arranged array of (3, 0)-CNTs, (9, 0)-CNTs and (3, 0)-(9, 0)-CNTs as indicated in Fig.S2 \cite{39}. The P6/mmm structure, with its relatively low energy as compared to previously proposed post-carbon-nanotube phases \cite{42, 43} may be a target for experimental synthesis. As such, we calculate the XRD of P6/mmm carbon and present it in Fig.S3 \cite{39} for future experimental comparison.

I$\overline{4}$3d is a topologically interesting cubic carbon phase with very high symmetry (No. 220). It is a complex structure containing 94 carbon atoms in its primitive cell (188 atoms in the conventional cubic cell). Only six of them are inequivalent. Such a large-sized and low-energy carbon structure presents a challenge to crystal structure prediction techniques \cite{35, 36, 37, 38}. It was first identified by the RG$^2$-code \cite{5} and then quickly rediscovered using the newest AIRSS \cite{35, 36} code. The carbon-rings in I$\overline{4}$3d are only 5-fold, 6-fold and 7-fold, which are different from the previously proposed sp$^3$ cubic carbons containing 3-fold, 4-fold, 6-fold and 8-fold carbon-rings. We find no potential pathways for transforming graphite and CNTs to I$\overline{4}$3d due to its complex network. However, in view of its stability, we might expect it to be discovered in the aerolite \cite{45} or synthesized in explosive shock experiments \cite{46}. The simulated XRD of I$\overline{4}$3d is also presented in Fig.S3 \cite{39}.

In Fig.~\ref{fig2}, we present a scatter plot of the PBE-based relative total energies (eV/atom) against equilibrium volume ($\AA^3$) of the carbon Pbam-32, P6/mmm, I$\overline{4}$3d and the previously proposed sp$^3$ carbons in the past decades, including those deposited in SACADA \cite{4} and those uncovered in our recent stochastic search \cite{5}. Pbam-32, P6/mmm and I$\overline{4}$3d show relatively low energies for the carbon family. As summarized in Tab.~\ref{tab-1} and Tab. S2, we can see that the total energies of the carbon-version Pbam-32, P6/mmm and I$\overline{4}$3d relative to cd are 96, 131 and 124 meV/atom, respectively, which indicate that they possess good stabilities comparable to the previously proposed Pbam-24 (69 meV/atom) \cite{16}, P41212 (135 meV/atom) \cite{16}, P42/ncm (108 meV/atom) \cite{15}, P6522 (109 meV/atom) \cite{14}, clat46 (106 meV/atom) \cite{17} and Clat34 (73 meV/atom) \cite{17}. In particular, I$\overline{4}$3d is noted as the third most stable cubic phase among those previously proposed, following clat34 (73 meV/atom) and clat46 (106 meV/atom). It is more stable than the experimentally synthesized T-carbon (1173 meV/atom) \cite{19, 20} and C22 (323 meV/atom) \cite{10, 46}. As show in Fig.S4 (a), the PBE-based enthalpy-pressure relations for Pbam-32, P6/mmm and I$\overline{4}$3d as carbon show that they keep good stability at high pressure in comparison with some well-known carbon phases.

The energetic stabilities of Pbam-32, P6/mmm and I$\overline{4}$3d as carbon indicate that they possess relatively higher probabilities to be synthesized. However, their dynamical stabilities and elastic stabilities should be determined. The phonon dispersions of Pbam-32, P6/mmm and I$\overline{4}$3d as carbon are computed (see methods in the supplemental material \cite{39}) and plotted in Fig.S5 (a). There are no imaginary frequencies in the phonon band structures, which indicate that Pbam-32, P6/mmm and I$\overline{4}$3d are dynamically stable for carbon. The elastic constants for carbon Pbam-32, P6/mmm and I$\overline{4}$3d are calculated and summarized in the supplemental material \cite{39} in Tab.S3 together with the corresponding bulk modulus, shear modulus, Young's modulus and Vicker's hardness\cite{39}. The elastic constants satisfy the mechanical stability criteria for corresponding crystal-systems, which indicate that Pbam-32, P6/mmm and I$\overline{4}$3d are mechanically stable as carbon. The calculated Vicker's hardness/bulk modulus of the carbon cd, Pbam-32, P6/mmm and I$\overline{4}$3d are 91.32/476.69, 83.58/456.95, 78.45/445.42 and 85.33/443.77 GPa, respectively. These hardness assessed using the empirical Chen's model\cite{47} suggest that the three new carbons are superhard materials \cite{39}.

\begin{figure*}
\center
\includegraphics[width=\textwidth]{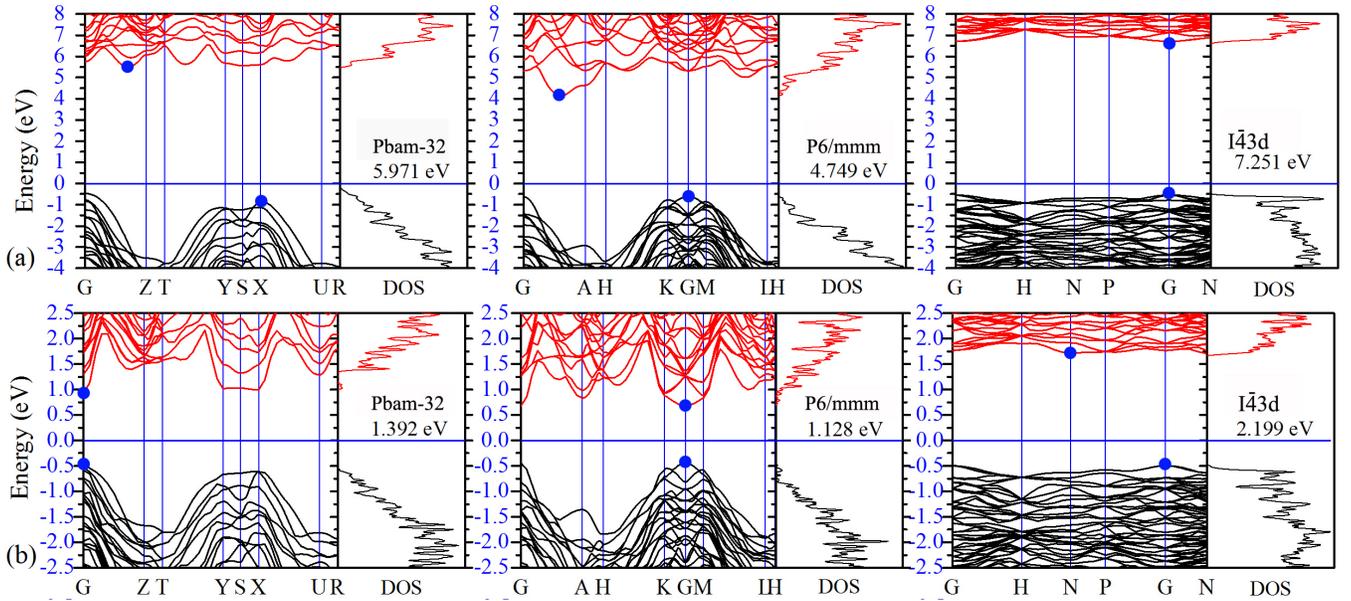}
\caption{The HSE06-based electronic band structures and density of states (DOS) of the newly discovered Pbam-32, P6/mmm and I$\overline{4}$3d structures as carbon (a) and silicon (b) at zero pressure.}\label{fig3}
\end{figure*}

The electronic properties of these three new structures of carbon are investigated by both the PBE and HSE06 methods. The calculated band-gaps are summarized in Tab.~\ref{tab-1} and Tab. S2 together with those of some selected tetrahedral structures. We can see that the PBE-based band gaps of the carbon Pbam-32, P6/mmm and I$\overline{4}$3d are 4.76, 3.55 and 5.91 eV. The band gap of 5.91 eV for I$\overline{4}$3d is about 0.7 eV larger than that of the previously proposed I-4 (5.25 eV) \cite{48}. It is about 0.93 eV larger than that of tP12 (4.98 eV) \cite{49}, therefore I$\overline{4}$3d is the widest-gap carbon allotrope known. To obtain more reliable band gaps, we then calculate the band structures of cd, I-4, tP12, Si24, Pbam-32, P6/mmm and I$\overline{4}$3d using HSE06. The HSE06-based band structures and density of state (DOS) of the carbon Pbam-32, P6/mmm and I$\overline{4}$3d are selected shown in Fig.~\ref{fig3} (a). As listed in Tab.~\ref{tab-1} and Tab. S2, the HSE06-based band gaps of cd, I-4 \cite{48, 28}, tP12\cite{49}, Si24\cite{32}, Pbam-32, P6/mmm and I$\overline{4}$3d are 5.32, 6.55, 6.27, 4.06, 5.97, 4.76 and 7.24 eV, respectively, which further confirmed that I$\overline{4}$3d is the widest-gap carbon allotrope.

In view of their tetrahedral topological atomic configurations, the crystalline Pbam-32, P6/mmm and I$\overline{4}$3d structures are likely in silicon, germanium and tin. After optimizing as silicon, germanium and tin, the topological characteristics of Pbam-32, P6/mmm and I$\overline{4}$3d are maintained. The lattice constants and bond lengths are correspondingly enlarged in comparison with those in carbon and the bond angles are well conserved. We have summarized all the structural information, including space groups, lattice constants, inequivalent atomic positions of Pbam-32, P6/mmm and I$\overline{4}$3d, as carbon, silicon, germanium and tin structures, in the supplemental tables \cite{39}.

The total energies, equilibrium volumes, PBE-based and HSE06-based band gaps of the Pbam-32, P6/mmm and I$\overline{4}$3d as silicon, germanium and tin are summarized in Tab.~\ref{tab-1} and Tab. S2. The total energies of the silicon-version Pbam-32, P6/mmm and I$\overline{4}$3d relative to cd silicon are 33, 46 and 39 meV/atom, respectively. These energies are comparable to those of the previously proposed Pbam-24 (29 meV/atom) \cite{16}, P41212 (39 meV/atom) \cite{16}, P42/ncm (44 meV/atom) \cite{15} and P6522 (53 meV/atom) \cite{14}. Especially, the silicon-version Pbam-32, P6/mmm and I$\overline{4}$3d are more stable than the experimentally viable cage structures Si24 (91 meV/atom) \cite{32}, clat34 (51 meV/atom) and clat46 (62 meV/atom), which are highly expected to be synthesized in future. Both germanium and tin Pbam-32, P6/mmm and I$\overline{4}$3d possess good energetic stabilities. For germanium, the total energies relative to cd germanium are 31, 46 and 35 meV/atom, respectively. For tin, the relative total energies are 26, 39 and 31 meV/atom, respectively. The plotted PBE-based enthalpy-pressure relations for these three new structures and some selected old structures as silicon (Fig.S4 (b)), germanium (Fig.S4 (c)) and tin (Fig.S4 (d)) show that Pbam-32, P6/mmm and I$\overline{4}$3d possess good stability at high pressure.

As shown in Fig.S5, the calculated phonon spectrums indicate that Pbam-32, P6/mmm and I$\overline{4}$3d are also dynamically stable for silicon, germanium and tin. The elastic constants summarized in the supplementary file for Pbam-32, P6/mmm and I$\overline{4}$3d as silicon (Tab.S4), germanium (Tab.S5) and tin (Tab.S6) also satisfy the corresponding mechanical stability criteria, indicating that they are also mechanically stable phases for silicon, germanium and tin. The Vicker's hardness assessed using Chen's empirical model \cite{47} summarized in Tab. S4, S5 and S6 suggest that silicon, germanium and tin with structures of Pbam-32, P6/mmm and I$\overline{4}$3d possess similar Vicker¡¯s hardness to those in cd structure.

The electronic properties of these three new structures as silicon, germanium and tin are also investigated by both the PBE and HSE06 methods. Their corresponding band gaps are summarized in Tab.~\ref{tab-1} and Tab. S2. The HSE06-based band structures and density of states of these three structures (Pbam-32, P6/mmm and I$\overline{4}$3d) for silicon are shown in Fig.~\ref{fig3}(b). As silicon, Pbam-32 possesses a PBE-based direct band gap of 0.81 eV and a HSE06-based direct band gap of 1.39 eV. The PBE-based gap of P6/mmm is 0.56 eV and its corresponding HSE06 gap is 1.13 eV. The HSE06-based band gaps of Pbam-32 and P6/mmm are considerably closer to the optimal value of 1.5 eV for photovoltaic applications, similar to that of the recently synthesized Si24 \cite{32}. The silicon I$\overline{4}$3d structure, has PBE and HSE06 indirect band gaps of 1.54 and 2.19 eV, respectively. The silicon I$\overline{4}$3d possesses the largest band gap among these selected structures.

The germanium Pbam-32 structure possesses a direct PBE band gap of 0.08 eV. This band gap is 0.66 eV in HSE06 as shown in Fig.S6 (c) \cite{39}. P6/mmm germanium is a direct band gap semiconductor (0.27 eV) with PBE-method. However, it becomes quasi-direct band gap semiconductor (0.77/0.88 eV) under HSE06. The I$\overline{4}$3d germanium is a quasi-direct band gap semiconductor with PBE-based and HSE06-based indirect band gaps of 0.85 and 1.41 eV, respectively. These indirect band gaps are slightly smaller than the corresponding direct band gaps of 1.06 and 1.58 eV. The HSE06 band structures and density of state of the tin Pbam-32, P6/mmm and I$\overline{4}$3d structures are provided in Fig.S6 (d) \cite{39}. The tin-version Pbam-32 and P6/mmm structures are metals within PBE, similar to the cd structure. However, they are narrow band gap semiconductors within HSE06, with gaps of 0.07 and 0.4 eV, respectively. The I$\overline{4}$3d tin structure is an indirect band gap semiconductor, its PBE-based band gap is 0.47 eV which will increases to be 0.86 eV in HSE06. These gaps indicate that tin-version I$\overline{4}$3d is a medium band gap semiconductor, similar to the previously proposed Si24 \cite{32} and I-4 \cite{48, 28} structures.

In summary, three tetrahedral networks (Pbam-32, P6/mmm and I$\overline{4}$3d) uncovered by a stochastic search strategy using the AIRSS and RG$^2$-codes, are investigated as carbon, silicon, germanium and tin structures through first-principles computations. The newly identified Pbam-32, P6/mmm and I$\overline{4}$3d structures contain 32, 36 and 94 atoms in their primitive cells, respectively. For carbon, their energies relative to cd are 96 meV/atom, 131 meV/atom and 112 meV/atom, respectively, showing good energetic stability. They are further confirmed to be dynamically and elastically stable superhard insulators, which are expected to be synthesized experimentally. I$\overline{4}$3d is a wide-gap semiconductor with a 7.25 eV gap, which is the widest know band gap of any carbon solid. These three new structures are also confirmed to be viable phases for silicon, germanium and tin, also showing remarkable energetic stability and positive dynamical and elastic stabilities. Their calculated electronic properties are suitable for semiconductor and photovoltaic applications.\\

This work is supported by the National Natural Science Foundation of China (Grant No. 11704319), the National Basic Research Program of China (2015CB921103), the Natural Science Foundation of Hunan Province, China (Grant No. 2016JJ3118) and the Program for Changjiang Scholars and Innovative Research Team in University (IRT13093). CJP is supported by the Royal Society through a Royal Society Wolfson Research Merit award and the EPSRC through grants EP/P022596/1 and EP/J010863/2.
\bibliographystyle{apsrev}

\end{document}